# Neutrino masses and mixing from hierarchy and symmetry

**Peter Kaus[a], Sydney Meshkov[b]**

[a] Physics Department, University of California, Riverside, CA 92521, USA
[b] California Institute of Technology, Pasadena, CA 91125, USA

**Abstract**

We construct a model that allows us to determine the three neutrino masses and the mass matrix directly from the experimental mass squared differences $\Delta_{atm}$ and $\Delta_{sol}$, using the assumptions of rational hierarchy ($m_1/m_2=m_2/m_3$) and of S3-S2 symmetry for the mixing matrix. We find that both the mass ratios and mixing angles are dominated by a parameter $\Lambda$. For the mixing angles, $\Lambda=\sqrt{(1/6)}\approx 0.41$, is a Clebsch-Gordan coefficient. For the masses, the mass ratios depend on the experimental $\Delta_{atm}$ and $\Delta_{sol}$ and with most recent data, remarkably, we also obtain $\sqrt{(m_1/m_2)}=\sqrt{(m_2/m_3)}$ $=0.41 = \Lambda$. This possibly coincidental equality gives a simple mass matrix in the $\sin(\theta_{13})=0$ limit.. We find that with $\Delta_{sol}= 8.2 \times 10^{-5}$ eV$^2$, $m_1= 1.5 \times 10^{-3}$ eV, $m_2=9.2 \times 10^{-3}$ eV and $m_3=5.3(5.5) \times 10^{-2}$ eV for $\Delta_{atm} = 2.73\ (2.95) \times 10^{-3}$eV$^2$. We obtain the mass matrix M and evaluate it's elements numerically for the presently 'best fit' solution in the allowed range for $\sin(\theta_{13})$. We find that all matrix elements are smaller than 0.03 eV. The only candidates for double texture zeroes are $M_{ee}$ and $M_{e\tau}$ or $M_{e\mu}$ (with $\theta_{13} \to -\theta_{13}$). The maximum effective mass for neutrinoless $\beta\beta$ decay is $|m_{\beta\beta}|_{max} \approx 8 \times 10^{-3}$ eV.

## 1. Introduction

The twelve fermion masses of the Standard Model are, at present, arbitrary parameters. A Grand Unified Model might, in principle, establish some relations among them. Although a very promising approach exists [1], there is no generally accepted model that establishes such relations. One of the things that we can do, in the mean time, is to look for empirical relationships or patterns. One such pattern, that of the 'rational' hierarchy of quarks and charged leptons, is well confirmed. By rational we mean that mass-ratios of members of a family are very close to powers of a parameter $\lambda$ [2]. For example, $m_b:m_s:m_d \approx 1:\lambda^2:\lambda^4$. This parameter also dominates the symmetry breaking exhibited by the mixing angles of the unitary matrix, which gives the flavor states as linear combinations of the mass eigenstates.

Mass patterns for neutrinos appear to be quite different from those of the charged fermions.. The information for neutrinos comes mainly from solar and atmospheric neutrino oscillations [3], [4].

$\Delta_{sol} = |m^2_{\nu 2} - m^2_{\nu 1}| \approx 8.2^{+0.6}_{-0.5} \times 10^{-5}$ eV$^2$ and $\Delta_{atm} = |m^2_{\nu 3} - m^2_{\nu 2}| \approx 2.73^{+0.8}_{-1.0} \times 10^{-3}$ eV$^2$, each at 90% CL.

In the following we determine the neutrino masses by proposing that, they too, follow a rational hierarchy and we determine the mass matrix by imposing S3-S2 symmetry. We observe a new relation between the mixing angles and the mass ratios. The mixing angle $\theta_{13}$ is small. In the limit $\theta_{13}$ goes to zero, we impose S3-S2 symmetry on the mixing matrix to fix the remaining mixing angles $\theta_{23}$ and $\theta_{12}$. In a rational hierarchical model $m_2\approx\sqrt{\Delta_{sol}}$ and $m_3\approx\sqrt{(\Delta_{atm}+\Delta_{sol})}$. It then follows that $m_1$ must be small compared to $\Delta_{sol}$. Motivated in part by the observed numerical similarity of $s_{12}s_{23}$ and $(\Delta_{sol}/\Delta_{atm})^{(1/4)}$, we equate the Cabibbo angle, $\sqrt{(m_1/m_2)}= \sqrt{(m_2/m_3)}$, to $s_{12}s_{23}=s_{12}c_{23}$, which will be named $\Lambda$, similar in spirit, but not in magnitude to the Wolfenstein parameter, $\lambda$. With this identification, the masses and the mass matrix are totally determined by $\Delta_{sol}$ and $\Delta_{atm}$.

## 2. Symmetry and Hierarchy Lead to a Proposed New Parameter for Neutrino Mass Determination

The flavor states $\nu_e$,$\nu_\mu$ and $\nu_\tau$ are related to the mass eigenstates $\nu_1$, $\nu_2$ and $\nu_3$ by the unitary transformation U.

$$U=\begin{pmatrix} c_{12}c_{13} & -s_{12}c_{13} & s_{13}e^{-i\partial} \\ s_{12}c_{23}+c_{12}s_{13}s_{23}e^{i\partial} & c_{12}c_{23}-s_{12}s_{13}s_{23}e^{i\partial} & -c_{13}s_{23} \\ s_{12}s_{23}-c_{12}s_{13}c_{23}e^{i\partial} & c_{12}s_{23}+s_{12}s_{13}c_{23}e^{i\partial} & c_{13}c_{23} \end{pmatrix} \quad (1)$$

There are two 'large' angles $\theta_{21}$ and $\theta_{23}$. Setting the small angle $\theta_{13}$, for which there is as yet no lower limit, equal to zero, we obtain $U_0$:

$$U_0=\begin{pmatrix} c_{12} & -s_{12} & 0 \\ s_{12}c_{23} & c_{12}c_{23} & -s_{23} \\ s_{12}s_{23} & c_{12}s_{23} & c_{23} \end{pmatrix} \quad (2)$$

The three columns of U are the three eigenvectors of the mass matrix in the $\theta_{13}$=0 limit. If $V_i$ is the ith column of U (i=1,2,3), then the mass matrix M is given by:

$$M=\sum_i m_i V_i V_i^T \quad (3)$$

where $m_i$ is the ith eigenvalue of M.

It was proposed more than 15 years ago that the 'mass gap' of the hierarchical pattern is associated with pairing forces in analogy with Cooper pairs in BCS theory and the mass matrix of the neutral pseudoscalar mesons [5]. In this limit, the mass matrix is 'democratic' [6] and when diagonalized gives rise to only one massive state, the coherent state. The 'democratic' vector $V_d$ is of particular interest here, where

$$V_d=\sqrt{(1/3)}\begin{bmatrix}1\\1\\1\end{bmatrix} \quad (4)$$

and

$$V_d V_d^{\dagger}=(1/3)\begin{pmatrix}1&1&1\\1&1&1\\1&1&1\end{pmatrix}, \quad (5)$$

the 'democratic' matrix. The vector $V_d$ was assigned to the heaviest mass, $m_3$, with pairing forces creating the mass gap in mind. The masses $m_2$ and $m_1$ were thought to be generated through a breaking of this $S_3$ symmetry, $S_3 \rightarrow S_2 \rightarrow S_1$ [5], [7].

However, the smallness (or vanishing) of $\theta_{13}$ makes the BCS type mass gap interpretation for $m_3$ untenable in the neutrino case. In contrast to the BCS case, because of the vanishing of $U_{e3}$ in $U_o$ (Eq. (2)), $m_3$ is a coherent mixture of $m_{\nu\mu}$ and $m_{\nu\tau}$, if $\theta_{23}=\pi/4$. Thus, we now have $S_2$ symmetry for $m_3$ and reserve $S_3$ symmetry for $m_2$. $U_0$ is now completely determined. This assignment of $V_d$ as the eigenvector for $m_2$ has lately received considerable attention in the literature [8].

If rational hierarchy is our aim, then we see from the definitions of $\Delta_{sol}$ and $\Delta_{atm}$, that if $m_1 << \Delta_{sol}$ then the mass ratios $\surd(m_1/m_2) \approx \surd(m_2/m_3) \approx (\Delta_{sol}/\Delta_{atm})^{(1/4)}$ are implied. In fact, the 90% confidence limits on $\Delta_{sol}$ and $\Delta_{atm}$ with present data, imply that $0.39 \leq \surd(m_1/m_2) \approx \surd(m_2/m_3) \leq 0.46$.. For the 'best fit' we obtain $\surd(m_1/m_2) \approx \surd(m_2/m_3) \approx 0.41$. Considering that from S3 symmetry we have $s_{12}\, s_{23} = \Lambda = \surd(1/6) = 0.41$, we suggest that this rough equality is not a coincidence.

We now relate the second large mixing angle, $\theta_{12}$, , to the mass ratio $m_1/m_2$ by the relation:

$$-s_{12}\, s_{23} \equiv \Lambda = \surd(m_1/m_2)\ . \tag{6}$$

This association of the mixing angles with the mass ratios was suggested by us earlier on phenomenological grounds [7], because both $s_{12}s_{23}$ and $(\Delta_{sol}/\Delta_{atm})^{(1/4)}$ are about the same, approximately equal to 0.4. We propose it here as a `natural' pattern.

Considering $s_{12}$ a small parameter for the moment (it is not), we get to first order in $s_{12}$ ($c_{12}=1$) the matrix $u_0$:

$$u_0 = \begin{pmatrix} 1 & \sqrt{2}\,\Lambda & 0 \\ -\Lambda & \sqrt{1/2} & -\sqrt{1/2} \\ -\Lambda & \sqrt{1/2} & \sqrt{1/2} \end{pmatrix} \tag{7a}$$

or more suggestively:

$$u_0 = \begin{pmatrix} 1 & \Lambda & \Lambda \\ -\Lambda & 1 & 0 \\ -\Lambda & 0 & 1 \end{pmatrix} \begin{pmatrix} 1 & 0 & 0 \\ 0 & \sqrt{1/2} & -\sqrt{1/2} \\ 0 & \sqrt{1/2} & \sqrt{1/2} \end{pmatrix} \tag{7b}$$

This shows the dynamic role assigned to $\theta_{12}$ by the assumption (6) and why we may consider it as 'natural'.

Restoring $c_{12}$ and full unitarity we have for $U_0$:

$$U_0 = \begin{pmatrix} \sqrt{(1-2\Lambda^2)} & \sqrt{2}\,\Lambda & 0 \\ -\Lambda & \sqrt{1/2}\sqrt{(1-2\Lambda^2)} & -\sqrt{1/2} \\ -\Lambda & \sqrt{1/2}\sqrt{(1-2\Lambda^2)} & \sqrt{1/2} \end{pmatrix} \tag{8}$$

Imposing $S_3$ symmetry (democracy) for the vector $V_2$ implies $U_{e2}=U_{\mu 2}=U_{\tau 2}$ or $\surd 2\,\Lambda = \surd(1/2)\surd(1-2\Lambda^2)$, so that

$$U_0 = \begin{pmatrix} 2\Lambda & \sqrt{2}\,\Lambda & 0 \\ -\Lambda & \sqrt{2}\,\Lambda & -\sqrt{1/2} \\ -\Lambda & \sqrt{2}\,\Lambda & \sqrt{1/2} \end{pmatrix} \tag{9}$$

By normalization, it follows that

$$-s_{12}\, s_{23} = \Lambda = \sqrt{(1/6)} \tag{10}$$

Of course $\sqrt{(1/6)}$ is not a capricious number, inasmuch as along with $\sqrt{(1/2)}$ it is a Clebsch-Gordan Coefficient. What is a capricious notion is that it also is numerically equal to $\sqrt{(m_1/m_2)}$, which follows from the assumption of rational hierarchy, $\sqrt{(m_1/m_2)} = \sqrt{(m_2/m_3)}$**,** as shown below.

For the mass ratios, we have

$$m_1/\,m_2 = m_1/\sqrt{(\Delta_{sol} + m_1^{\,2})} \tag{11a}$$

and

$$m_2/m_3 = \sqrt{(\Delta_{sol} + m_1^{\,2})}/\sqrt{(\,\Delta_{atm} + \,\Delta_{sol} + m_1^{\,2})} \tag{11b}$$

Setting $m_1/\,m_2 = m_2/m_3$, and solving for $m_1$ using present data ($\Delta_{sol} = 8.2 \times 10^{-5}\,eV^2$ and $\Delta_{atm} = 2.75 \times 10^{-3}\,eV^2$) , we get $m_1 = 1.5 \times 10^{-3}$ eV and $\sqrt{(m_1/m_2)} = \sqrt{(m_2/m_3)} = 0.41 = \sqrt{(1/6)}$,.

It is, of course entirely possible that it is a coincidence that $s_{12}s_{23} \approx (\Delta_{sol}/\Delta_{atm})^{(1/4)}$ and that both are approximately equal to $\sqrt{(1/6)}$, which is the value demanded by S3-S2 symmetry, but we make this equality the basis of the present model. Hence Eq (10).

We now have $-\sin(\theta_{23}) = \cos(\theta_{23}) = \sqrt{(1/2)}$ and $-\sin(\theta_{12}) = \sqrt{(1/3)} = \sqrt{2}\,\Lambda$, so that $\tan^2(\theta_{23}) = 1$ and $\tan^2(\theta_{12}) = 1/2$**.**

The hierarchy indicated here is not very strong, $m_2 \approx \Lambda^2\, m_3 = (1/6)\, m_3$, so $\Lambda$ should not be used as an expansion parameter. In fact, the situation is very different from the quark sectors. There, the possible $S_3$-$S_2$ symmetry is presumably the same for the d and u sectors and does not appear in the $V_{ckm} = U_d^{\dagger}\, U_u$, which is then just 1. Only the symmetry breaking terms, dominated by powers of $\lambda \approx 0.23$ are seen and the underlying symmetry, if it exists, is obscured in the resulting Wolfenstein representation. The mixing angles can be large or small, depending on the assumed flavor basis. In the present model, on the other hand, $\Lambda$ is intrinsic to the symmetry and must be $\sqrt{(1/6)} \approx 0.41$. An even weaker hierarchy has been proposed by Xing [9] , where U and M are 'expanded' in terms of $\Lambda = U_{\mu 3} \approx \sin(\theta_{23}) \approx 0.7 \approx \sqrt{(m_2/m_3)}$.

## 3. The neutrino mass spectrum

Assuming the normal ordering of masses, $m_1^2 < m_2^2 < m_3^2$ , we have two equations, for $m_2^2$ and $m_3^2$ in terms of the experimentally observed mass squared differences, $\Delta_{sol}$ and $\Delta_{atm}$.

$$m_2^2 = \Delta_{sol} + m_1^2 \qquad (12)$$

$$m_3^2 = \Delta_{atm} + \Delta_{sol} + m_1^2 \qquad (13)$$

A mass scale is provided by a third equation , which relates $m_l$ and $m_2$ (see discussion above),

$$\sqrt{(m_1/m_2)} = \sqrt{(1/6)} = \Lambda \qquad (14)$$

Without loss of generality, but with an eye towards 'rational' hierarchy, we now represent the masses $m_1$, $m_2$, $m_3$ in terms of parameters $\Lambda$, $m_3$ and $\mu$, where $\mu$, measures the deviation from rational hierarchy.

$$M_{diag} = \begin{pmatrix} m_1 & 0 & 0 \\ 0 & \sqrt{\Delta_{sol} + m_1^2} & 0 \\ 0 & 0 & \sqrt{\Delta_{atm} + \Delta_{sol} + m_1^2} \end{pmatrix} = m_3 \begin{pmatrix} \mu\Lambda^4 & 0 & 0 \\ 0 & \mu\Lambda^2 & 0 \\ 0 & 0 & 1 \end{pmatrix} \qquad (15)$$

Thus $m_2/m_3 = \mu\Lambda^2$ and $m_1/m_2 = \Lambda^2$. 'Perfect' rational hierarchy would mean $\mu$=1. The data for $\Delta_{sol}$ and $\Delta_{atm}$ considerably restrict the possible solutions.

In Fig. 1 we display

$$\Delta_{atm} = \Delta_{sol}\left(\frac{\Lambda^4\mu^2 - 1}{\Lambda^4\mu^2(\Lambda^4 - 1)}\right) \qquad (16)$$

which follows from the representation (15), to see the range, if any, of solutions consistent with the experimental range of $\Delta_{sol}$ and $\Delta_{atm}$ . For clarity of the figure we chose a rectangle slightly smaller than the 2$\sigma$ limits of Araki et.al., [3], and of Nakaya [4],

$$7.7 \times 10^{-5}\ eV^2 < \Delta_{sol} < 8.8 \times 10^{-5}\ eV^2 \text{ and } 1.9 \times 10^{-3}\ eV^2 < \Delta_{atm} < 3.0 \times 10^{-3}.eV^2$$

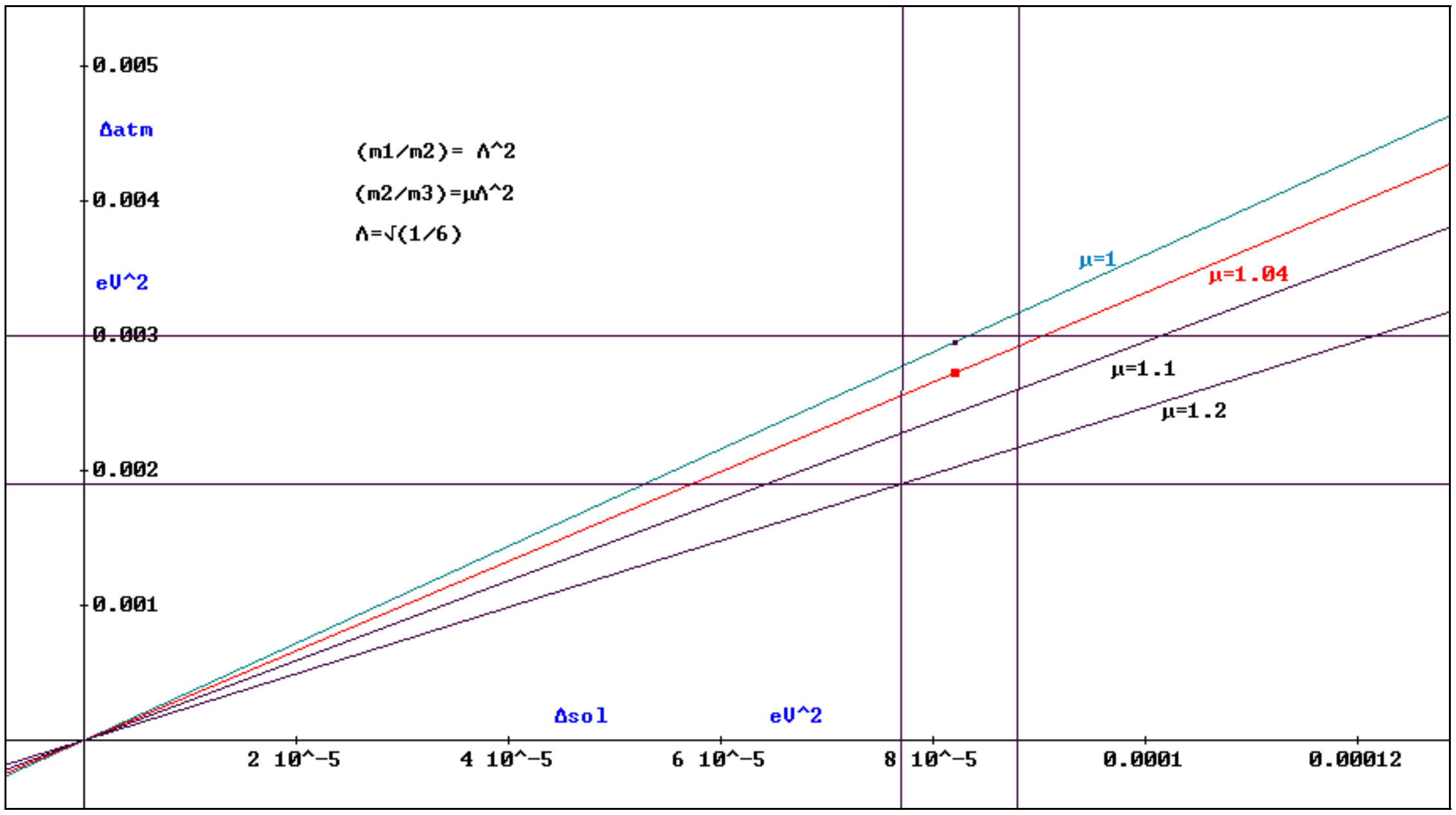

***Fig.1*** ***$\Delta_{atm}$ vs. $\Delta_{sol}$ (Eq. 16 with $\Lambda^2$=1/6) for various values of μ . ($\mu m_1/m_2 = m_2/m_3 = \mu\Lambda^2$). Acceptable solutions are within the (slightly arbitrary) rectangle 1.9 x $10^{-3}$ $eV^2$ < $\Delta_{atm}$ < 3 x $10^{-3}$ $eV^2$ and 7.7 x $10^{-5}$ $eV^2$<$\Delta_{sol}$<8.8 x $10^{-5}$ $eV^2$. Two solutions are marked. They correspond to $\Delta_{sol}$ = 8.2 x $10^{-5}$ $eV^2$, with μ=1 (rational hierarchy) and μ=1.04 (best fit),[4]).***

With $\Delta_{sol}$ and $\Delta_{atm}$ given, all mass values are fixed. It is gratifying to have μ ≈ 1 to be squarely in the acceptable data range, because values of μ much different from unity depart from the spirit of rational hierarchy. It is clear from Fig.1 that μ ranges from 0.9-1.2 with $\Lambda=\sqrt{(m_1/m_2)}=\sqrt{(1/6)}$. For the 'best fit' values, $\Delta_{sol}$=8.2 x $10^{-5}$ and $\Delta_{atm}$=2.73 x $10^{-3}$ $eV^2$, we have μ=1.04. This demonstrates the consistency of the model (Eq. 6) with the notion of rational hierarchy and the oscillation data.

Equations (15) can be solved to give the masses and the ratio parameter μ, ($\mu\, m_1/m_2 = m_2/m_3 = \mu\Lambda^2$), as functions of $\Delta_{atm}$ and $\Delta_{sol}$.

$$m_1^{\,2} = \Delta_{sol}\frac{\Lambda^4}{(1-\Lambda^4)} \tag{17a}$$

$$m_2^{\,2} = \Delta_{sol}\frac{1}{(1-\Lambda^4)} \tag{17b}$$

$$m_3^{\,2} = \Delta_{atm} + \Delta_{sol}\frac{1}{(1-\Lambda^4)} \tag{17c}$$

$$\mu^2 = \Delta_{sol}\frac{1}{\Lambda^4(\Delta_{atm}(1-\Lambda^4)+\Delta_{sol})} \tag{17d}$$

The equations (17) are valid independent of the choice of Λ. Substituting $\Lambda=\sqrt{(1/6)}$ from Eq. (10), we look at the properties of the two marked solutions of Fig. 1. Both have $\Delta_{sol}$=8.2 x $10^{-5}$ $eV^2$ (best fit) and differ only in that for the first solution we chose μ=1 (rational hierarchy) and let Eq. (16) determine $\Delta_{atm}$, while for the second, we take $\Delta_{atm}$=2.73 x $10^{-3}$ $eV^2$ (best fit) and let μ be determined by Eq. (17d).

Since $m_1$ and $m_2$ depend only on $\Lambda$ and $\Delta_{sol}$, these masses are the same for both solutions; $m_1 = 1.53 \times 10^{-3}$ eV and $m_2 = 9.18 \times 10^{-3}$ eV.

Solution-1: 'rational hierarchy'. Input, $\mu$=1: $\Delta_{atm} = 2.95 \times 10^{-3}$ eV$^2$ $m_3 = 5.51 \times 10^{-2}$ eV
Solution-2: 'best fit' . Input, $\Delta_{atm} = 2.73 \times 10^{-3}$ eV$^2$ **:** $\mu = 1.04$ $m_3 = 5.3 \times 10^{-2}$ eV.

Both solutions have the property $m_2 = 6\, m_1$ and $\mu m_3 = 6\, m_2$ with $\mu$=1 and 1.04 respectively. $\Delta_{atm} = |m_3^2 - m_2^2|$ and $\Delta_{sol} = |m_2^2 - m_1^2|$ are within the acceptable experimental limits. All masses listed are absolute values.

**4. Elements of the Mass Matrix and their Properties**

The Mass matrix M is given by

$$M = U\, M_d\, U^T \tag{18}$$

where U is given by (1) and

$$M_d = \begin{pmatrix} m_1 e^{i\alpha_1} & 0 & 0 \\ 0 & m_2 e^{i\alpha_2} & 0 \\ 0 & 0 & m_3 \end{pmatrix} \tag{19}$$

The phases $\alpha_1$ and $\alpha_2$ are the Majorana phases. To get the matrix elements of M, listed below, we take $\alpha_1$ to be 0, $\alpha_2$ to be $\pi$ and $\delta$ is the CP violating Dirac phase.

In the absence of symmetry breaking terms, $\sin(\theta_{13}) = 0$ , we obtain the simple mass matrix:

$$M = m_3 \begin{pmatrix} 4\Lambda^6 - 2\Lambda^4 & -2\Lambda^6 - 2\Lambda^4 & -2\Lambda^6 - 2\Lambda^4 \\ -2\Lambda^6 - 2\Lambda^4 & \Lambda^6 - 2\Lambda^4 + \frac{1}{2} & \Lambda^6 - 2\Lambda^4 - \frac{1}{2} \\ -2\Lambda^6 - 2\Lambda^4 & \Lambda^6 - 2\Lambda^4 - \frac{1}{2} & \Lambda^6 - 2\Lambda^4 + \frac{1}{2} \end{pmatrix} \tag{20}$$

where $\Lambda = \sqrt{(1/6)}$ and $m_3 = \sqrt{\left(\Delta_{atm} + \Delta_{sol} \frac{1}{(1-\Lambda^4)}\right)} = 5.5 \times 10^{-3}$ eV$^2$

The elements of M are then given by:

$$M_{ee} = 1/3\,(2m_1 - m_2) + s_{13}^2\,[\,1/3\,(m_1 + m_2) - m_1 + e^{-2i\delta}\,m_3] \tag{21a}$$

$$M_{e\mu} = M_{\mu e} = 1/\sqrt{2}\;\; c_{13}\,[\,-\sqrt{2/3}\,(m_1 + m_2) + s_{13}\,\{\,e^{\,i\delta}\,[\,m_1 - 1/3(m_1 + m_2)] - e^{-i\delta}\,m_3\}] \tag{21b}$$

$$M_{e\tau} = M_{\tau e} = 1/\sqrt{2}\;\; c_{13}\,[\,-\sqrt{2/3}\,(m_1 + m_2) - s_{13}\,\{\,e^{\,i\delta}\,[\,m_1 - 1/3(m_1 + m_2)] - e^{-i\delta}\,m_3\}] \tag{21c}$$

$$M_{\mu\mu} = 1/6\,(m_1 - 2m_2 + 3m_3) - 1/6\,s_{13}\,[e^{\,i\delta}\,2\sqrt{2}\,(m_1 + m_2) - s_{13}\{\,e^{2i\delta}(2m_1 - m_2) - 3\,m_3\}] \tag{21d}$$

$$M_{\mu\tau} = M_{\tau\mu} = 1/6\,(m_1 - 2m_2 - 3m_3) - 1/6[\,s_{13}^2\{\,e^{2i\delta}(2m_1 - m_2) - 3\,m_3\}] \tag{21e}$$

$$M_{\tau\tau} = 1/6\,(m_1 - 2m_2 + 3m_3) - 1/6\,s_{13}[-e^{\,i\delta}\,2\sqrt{2}\,(m_1 + m_2) - s_{13}\{\,e^{2i\delta}(2m_1 - m_2) - 3\,m_3\}] \tag{21f}$$

Fig.2, below, shows the elements of M for Solution-2 as functions of $\sin(\theta_{13})$ with $\delta$=0. The maximum allowed $\sin(\theta_{13}) \approx 0.25$.

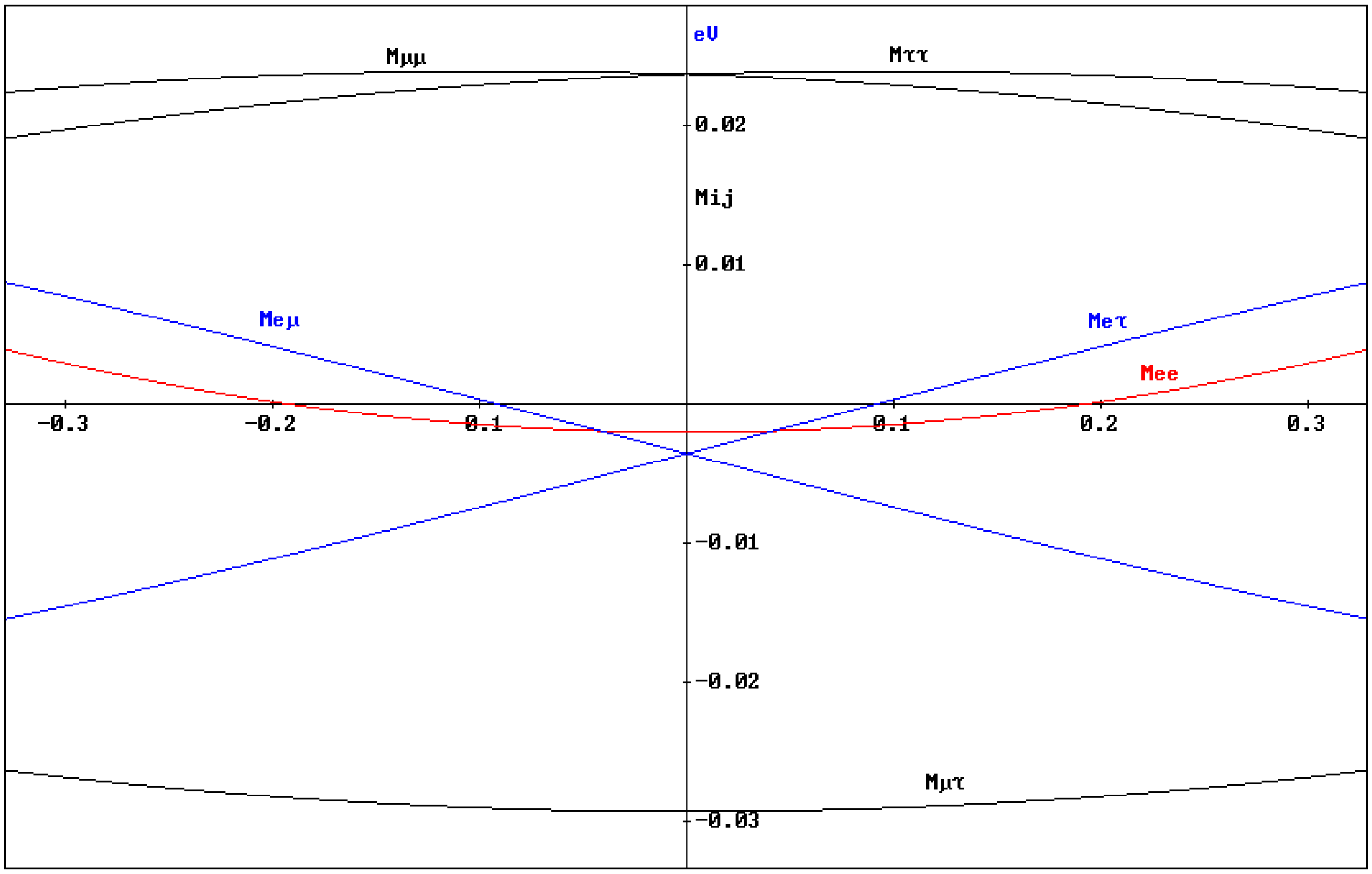

***Fig.2 Elements of the Mass matrix M as functions of $\sin(\theta_{13})$, with $\delta$=0. The masses are from Solution-2, the 'best fit' solution, $\mu$ =1.04, , $\Delta_{sol}$=8.2 x $10^{-5}$ eV$^2$ , $\Delta_{atm}$=2.73 x $10^{-3}$ eV$^2$, and $m_2\, e^{i\alpha_2}$= -9.18 x $10^{-3}$eV. All elements are smaller than 0.03 eV.***

As may be seen from Fig. 2, the only candidates for double texture zeroes [10] are $M_{ee}$ and $M_{e\tau}$ or $M_{e\mu}$ (with $\theta_{13} \rightarrow -\theta_{13}$). A double texture zero could be obtained with a moderate change in $\Delta_{sol}$ and $\Delta_{atm}$ [11], but not within their current experimentally acceptable limits. In addition, rational hierarchy would be badly violated. Consequently, we do not pursue this subject further.

The phase, $\delta$, of the mixing matrix, U, has a serious effect for the mass matrix for the matrix elements $M_{e\mu}$ and $M_{e\tau}$, because for these elements the real part vanishes in the allowed range for $\theta_{13}$., $\sin\theta_{13} \leq 0.25$ [12 ].

Fig.3 shows $|M_{e\tau}|$ vs. sin($\theta_{13}$) for various values of δ. The values for $|M_{e\tau}|$ and $|M_{\tau e}|$ are the same as those for $|M_{e\mu}|$ and $|M_{\mu e}|$, with $\theta_{13} \rightarrow -\theta_{13}$.

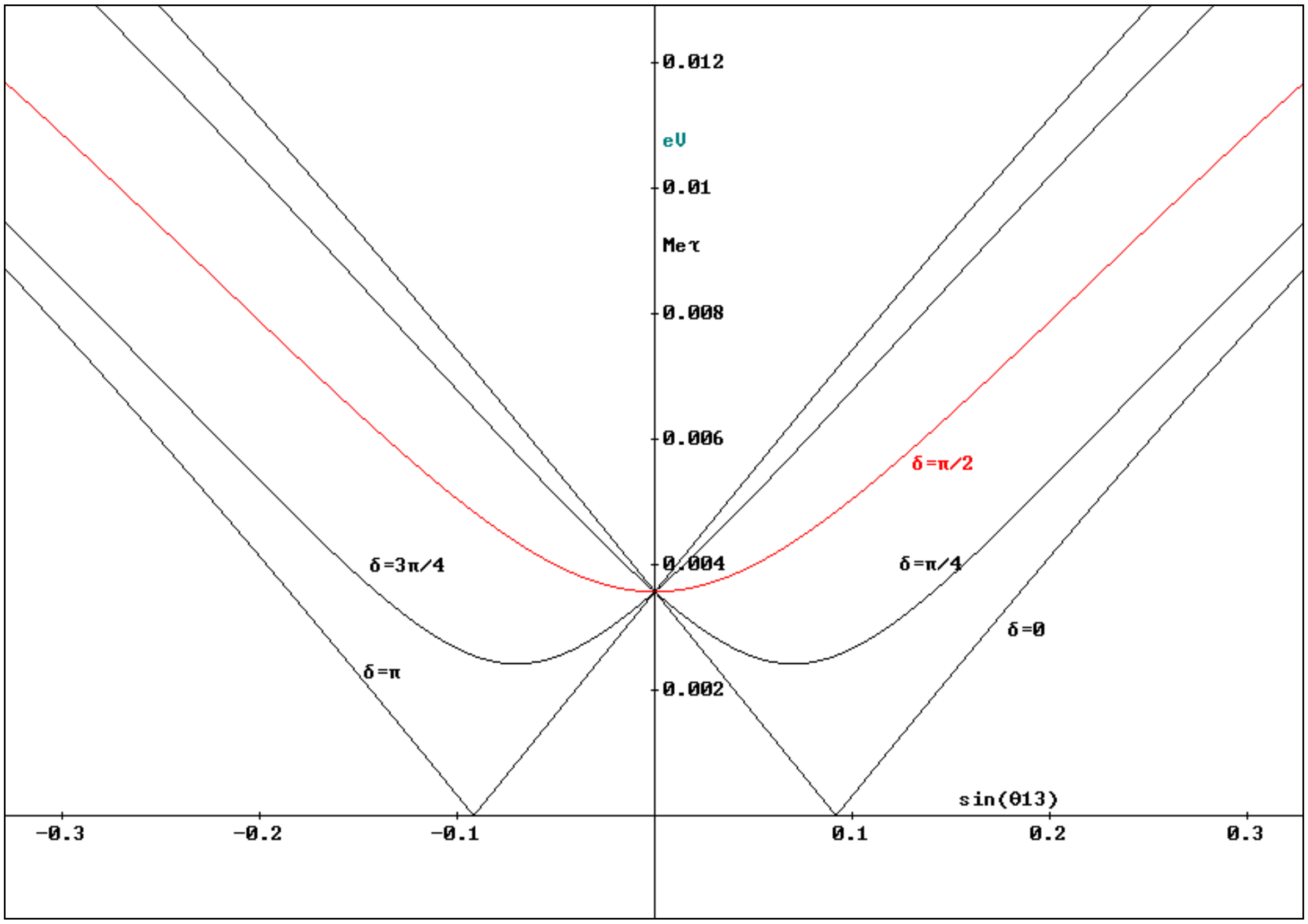

***Fig. 3*** ***$|M_{e\tau}|$ vs. sin($\theta_{13}$) for various values of δ, 0≤δ≤π in steps of π/4 for Solution 2.***

The effective mass for neutrinoless ββ decay is

$$|m_{\beta\beta}| = |\ (2/3)c_{13}^2\ m_1\ e^{i\varphi_1} + (1/3)\ c_{13}^2\ m_2\ e^{i\varphi_2} + s_{13}^2\ m_3\ | \tag{22}$$

where $\varphi_{1,2} = \alpha_{1,2} + 2\delta$ . To obtain $|m_{\beta\beta}|_{max}$ we set $\varphi_1$ and $\varphi_2 = 0$. Using $m_1 = 1.5\ \times 10^{-3}$ eV, $m_2 = 9.2\ \times 10^{-3}$ eV and $m_3 = 5.5\ \times 10^{-2}$ eV, we obtain $|m_{\beta\beta}|_{max} \approx 8\ 10^{-3}$ eV

**Conclusions**

We have applied 'rational' hierarchy, i.e. $m_1:m_2:m_3 = \Lambda^4:\Lambda^2:1$, to obtain the neutrino masses directly from the experimental mass squared differences, $\Delta_{atm}$ and $\Delta_{sol}$. The mass matrix was formulated with the assumption of S3-S2 symmetry for the mixing matrix.. Defining $-\sin(\theta_{12})\sin(\theta_{23}) = -\sin(\theta_{12})\cos(\theta_{23}) = \Lambda$, we find that Λ is the same both theoretically and derived from experimental data, i.e., $-s_{12}\ s_{23} = \sqrt{(m_1/m_2)} \equiv \Lambda = \sqrt{(1/6)}$. Consequently $m_1 \approx 1.5\ \times 10^{-3}$ eV and $m_2 \approx 9.2\ \times 10^{-3}$ eV. The largest mass, $m_3 \approx 5.5\ \times 10^{-2}$ eV$\approx\sqrt{(\Delta_{atm} + \Delta_{sol})}$. A study of the elements of the mass matrix, M, for our solution 2, that of the best fit solution, for the case δ = 0, shows that all of them are smaller than 0.03 eV. The phase, δ, of the mixing matrix U has a serious effect for the mass matrix for the matrix elements $M_{e\mu}$ and $M_{e\tau}$, because for these elements the real part vanishes in the allowed range for $\theta_{13}$., sin $\theta_{13}$ ≤ 0.25 . Their dependence on $s_{13}$ for various values of δ is shown explicitly. We find that the maximum effective mass for neutrinoless ββ decay is, $|m_{\beta\beta}|_{max} \approx 8\ \times 10^{-3}$ eV.

**Acknowledgements**

We thank Gabriela Barenboim, Douglas Michael, Pierre Ramond, and Ina Sarcevic for useful and informative discussions, We also are grateful for the support of the Caltech High Energy Theory Group and the U. S. Department of Energy in the preparation of this manuscript. We thank the Aspen Center for Physics for its kind hospitality.